\documentclass[aps,pre,amsmath]{revtex4}
\usepackage{graphics,epsfig}
\usepackage{bm}

\begin{document}

\title{Grain boundary dynamics in stripe phases of non potential systems}
\author{Zhi-Feng Huang}
\affiliation{Department of Physics and Astronomy, Wayne State
University, Detroit, MI 48201}
\affiliation{McGill Institute for Advanced Materials, and Department of
Physics, McGill University, Montreal, QC H3A 2T8, Canada}
\author{Jorge Vi\~nals}
\affiliation{McGill Institute for Advanced Materials, and Department of
Physics, McGill University, Montreal, QC H3A 2T8, Canada}

\date{\today}

\begin{abstract}
We describe numerical solutions of two non potential models of pattern
formation in nonequilibrium systems to address the motion and decay of
grain boundaries separating domains of stripe configurations of different
orientations. We first address wavenumber selection because of the boundary,
and possible decay modes when the periodicity of the stripe phases is
different from the selected wavenumber for a stationary boundary. We discuss
several decay modes including long wavelength undulations of the moving
boundary as well as the formation of localized  
defects and their subsequent motion. We find three different regimes as a
function of the distance to the stripe phase threshold and initial wavenumber, 
and then correlate these findings with domain morphology during domain coarsening
in a large aspect ratio configuration.
\end{abstract}

\pacs{47.54.-r, 05.45.-a, 47.20.Bp, 89.75.Kd}

\maketitle

\section{Introduction}

A stripe configuration is a particular example of a modulated configuration
that exhibits 
structural periodicity along only one spatial direction. In thermodynamic
equilibrium, these phases have been widely observed in a variety of systems
\cite{re:seul95}, including, for example, microphase separated block copolymers
\cite{re:bates99}. Outside of equilibrium, stripe patterns are also widespread,
including Rayleigh-B\'enard convection in fluids \cite{re:cross93}, or
electroconvection in nematic liquid crystals \cite{re:purvis01}. When the
lateral size of the system of interest is large 
compared with the wavelength of the modulation (large aspect ratio limit), a
stripe configuration in either class of systems is usually characterized by a
transient multidomain and defected 
configuration with only local order. Domain coarsening is
believed to be controlled by the motion of existing topological defects in the
configuration such as grain boundaries, dislocations, and disclinations. In
this paper we focus on non potential (or no variational) model systems and
investigate the processes of defect formation and motion near grain
boundaries, and show that there exist several regimes in parameter space
that are characterized by qualitatively different coarsening behavior.

A major topic of interest in domain coarsening of stripe patterns is the
growth law for the characteristic linear scale of the configuration. This has
been addressed both in theoretical studies
\cite{re:elder92,re:cross95a,re:boyer01,re:boyer01b,re:paul04} and in
experiments in nematics \cite{re:purvis01,re:kamaga04} and block copolymers
\cite{re:harrison00b}. This characteristic length scale $R(t)$ is expected to
obey a power law growth in time: $R(t) \sim t^{x}$, with a value of the coarsening
exponent $x$ which ranges from $1/5$ to $1/2$ in the studies to date. The
actual value appears to depend on 
$\epsilon$ (the dimensionless distance from the onset of the stripe phase)
\cite{re:boyer01b,re:boyer02}, thermal noise \cite{re:elder92}, and the
type of length scale under study \cite{re:cross95a,re:paul04}. Also, domain
morphologies as well as the value of $x$ have been found to be different in
potential systems (in which the evolution is solely driven by minimization of
a potential) than in the non potential case \cite{re:cross95a}.
Numerical studies of the Swift-Hohenberg model of Rayleigh-B\'enard convection
(a potential system) suggest that $x=1/3$ in the limit $\epsilon \ll 1$
as a result of grain boundary motion\cite{re:boyer01}. 
As $\epsilon$ increases, the value of $x$ is seen to
decrease from $1/3$, fact that was argued to be a manifestation of defect
pinning \cite{re:boyer02}. On the other hand, the generalized Swift-Hohenberg
model has been studied as an example of a non potential model
\cite{re:cross95a}, and it has been shown that the evolution of stripe
patterns is asymptotically dominated by the motion of isolated
dislocations. Furthermore, recent results from a direct solution of the
Boussinesq equations in a Rayleigh-B\'enard configuration suggest that
multiple length scales should be introduced in the case of non potential
systems \cite{re:paul04}.

\begin{figure}
\includegraphics[width=1.8in]{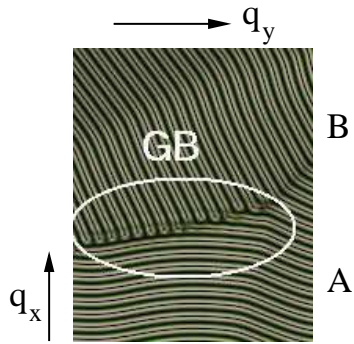}
\caption{A $90^{\circ}$ tilt grain boundary (GB) configuration of a stripe
  phase, with wave numbers $q_x$ and $q_y$ for two domains A and B of mutually
  perpendicular stripe orientations.}
\label{fig:GB}
\end{figure}

Despite a significant body of research on domain coarsening of modulated
patterns, the mechanisms responsible for the observed properties and the
values of the coarsening exponents, especially for non potential systems, are
not yet understood. 
We report in this paper our analysis of the relation between the selected
wavenumber at grain boundaries and the critical wavenumber, and consequences
on both early time defect formation and asymptotic coarsening. Our
starting point is a planar $90^{\circ}$ grain boundary separating two
stripe domains of mutually perpendicular orientations, as shown in
Fig. \ref{fig:GB}. We determine the selected wavevector for a stationary
boundary configuration ${\mathbf q^{s}}$ in two different non potential
models, and show that for other wavenumbers the mode of decay of the
configuration can be classified according to its wavenumber and the
dimensionless distance from threshold $\epsilon$. For small $\epsilon$, an
initially planar grain boundary separating two domains, one of which at least
has ${\mathbf q} \neq {\mathbf q^{s}}$, is seen to propagate, either as a
planar front, or by developing long wavelength undulations, depending on the
value of ${\mathbf q}$. For larger values of $\epsilon$ we find that the grain
boundary configuration decays through the formation of disclinations around the
boundary and their subsequent motion. We then correlate our findings with the
observed domain coarsening morphology of configurations obtained from
initially uniform but unstable states. The rationale for the comparison is the
observation that in an extended system an early time transient configuration
following the decay of an unstable state will comprise a distribution of
locally ordered domains with wavenumber approximately equal to the critical
wavenumber for instability $q_{0}$, but with different orientations. Depending
on whether $q_{0}$ is close to $q^{s}$ or not, the resulting boundaries will
be locally close to being stationary, or quickly decay leaving behind
different types of defects. The ensuing domain coarsening is shown to be
different depending on the (near) stability of grain boundaries that
spontaneously form in the early stages following the decay of the initially
unstable configuration. 

Our study is based on a numerical solution of two different non potential
extensions of the Swift-Hohenberg model of Rayleigh-B\'enard convection
\cite{re:cross93}, as described in Sec. \ref{sec:models}.  
One of them is made non potential on account of the specific
choice of nonlinearity, whereas the other allows for mean flows representing
the effect of vertical vorticity encountered in Rayleigh-B\'enard convection
in small Prandtl number fluids \cite{re:greenside85}. They are described
in Sections \ref{sec:model1} and \ref{sec:model2} respectively. Although wave
number selection for a stationary grain boundary is different in
both models, similar properties concerning grain boundary motion and
decay have been found. They correlate with two types of domain coarsening
behavior, either grain boundary dominated or disclination/dislocation
dominated, as deduced from the results on grain boundary decay that will be
discussed in Sec. \ref{sec:discussion}.

\section{Non potential model equations}
\label{sec:models}

The model systems we have investigated are two dimensional
generalized Swift-Hohenberg equations which have been used as
a phenomenological description of stripe patterns
\cite{re:greenside85,re:cross93}: 
\begin{equation}
\partial_t \psi + {\bm U} \cdot {\bm \nabla} \psi
= \epsilon \psi - (\nabla^2 + q_0^2)^2 \psi
-g \psi^3 + 3 (1-g) |\nabla \psi|^2 \nabla^2 \psi,
\label{eq:S-H}
\end{equation}
where $\psi(x,y,t)$ is a dimensionless real order parameter field, and the control
parameter $\epsilon$ is the distance from the threshold for instability of the
solution $\psi = 0$ (at $\epsilon = 0$). The wavenumber 
$q_0$ ($q_{0} = 1$ in the dimensionless units that we are using) is the
critical wave number for instability. An advection term with a drift velocity
${\bm U}(x,y,t)=(U_x,U_y)$ has been added to Eq. (\ref{eq:S-H}) to introduce mean
flow effects resulting from the coupling to vertical vorticity in the fluid
\cite{re:manneville83,re:greenside85}. The velocity  ${\bm U}(x,y,t)$ is given
in terms of the vorticity potential $\zeta(x,y,t)$:
\begin{equation}
{\bm U} = {\bm \nabla} \times (\zeta \hat{\bm z})
= \left ( \partial_y \zeta, -\partial_x \zeta \right ), 
\label{eq:U_drift}
\end{equation}
which satisfies,
\begin{equation}
\left [ \partial_t - \sigma (\nabla^2 - c^2) \right ] \nabla^2 \zeta
= g_m \left [ {\bm \nabla}(\nabla^2 \psi) \times {\bm \nabla} \psi
\right ] \cdot \hat{\bm z},  
\label{eq:mean_flow}
\end{equation}
with $\sigma$ being the Prandtl number of the fluid,
$c^2$ introduced to phenomenologically model the effect of no slip boundary
conditions at the top and bottom plates of the convection cell
\cite{re:manneville83}, and the coupling parameter $g_m$ chosen to be a
decreasing function of the Prandtl 
number. Following Ref. \onlinecite{re:greenside85} a Gaussian filtering
operator $F_{\gamma}$ is introduced on the r.h.s. of
Eq. (\ref{eq:mean_flow}) to reduce high wavevector contributions to
the vertical vorticity. In Fourier space, $F_{\gamma}$ acts by
multiplying the Fourier transform of the filtered function by a factor
\begin{equation}
\hat{F}_{\gamma_q} = e^{-\gamma^2 q^2},
\nonumber
\end{equation}
where $q^2=q_x^2+q_y^2$, with wavevector components $q_x$ and $q_y$,
and $\gamma$ represents the filtering radius.

Different types of model equations for describing stripe patterns have been
included in Eqs. (\ref{eq:S-H})--(\ref{eq:mean_flow}), depending on the values
of $g$ and $g_m$. For $g=1$ and $g_m =0$ (so that ${\bm U}=0$), the
nonlinearity in the model is given by the cubic term $\psi^3$ only, leading to the
original Swift-Hohenberg equation \cite{re:swift77}, which is a
potential model. A non potential model without mean flows follows from
$g=g_m=0$ (referred to as model 1 in this paper), which includes a nonlinear
term of the form $|\nabla \psi|^2 \nabla^2 \psi$. For $g=1$ and $g_m \neq 0$
(and hence ${\bm U} \neq 0$), we obtain the Swift-Hohenberg model supplemented
by mean flows, also a non potential system (model 2). 

The initial configuration considered in the first part of our study is a
$90^{\circ}$ tilt boundary separating two stripe domains, as shown in
Fig. \ref{fig:GB}. A stationary state is known to exist \cite{re:tesauro87} if
unique values of the wavenumbers $q_x$ (for the stripes parallel to the grain
boundary in domain A) and $q_y$ (for the stripes perpendicular to the boundary
in domain B) are selected in both domains. For the original Swift-Hohenberg
model, these wavenumbers are $q_{x}=q_{y}=q_{0}$. In our numerical study, we
consider instead a symmetric pair of grain boundaries along the $x$ direction
so that periodic boundary conditions can be adopted. The pseudospectral
algorithm described in Ref. \onlinecite{re:cross94b} has been used to numerically
solve the model equations, with a time step $\Delta t=0.2$ and a grid spacing 
$\Delta x = \Delta y=\lambda_0 /8$ (i.e., 8 grid points per stripe
wavelength $\lambda_0=2\pi/q_0$). Most of our results shown below correspond
to a system size $512 \times 512$ grid nodes, with spot checks with larger sizes
($2048 \times 2048$) to verify our results.

\section{Wavevector selection and grain boundary decay - Model~1}
\label{sec:model1}

\subsection{Wavenumber selection}

Model 1 includes the nonlinear term $|\nabla \psi|^2 \nabla^2 \psi$ but
does not have mean flow. The selected wave numbers ($q_x^s$ and $q_y^s$) for a
stationary grain boundary configuration have been already calculated by
Tesauro and Cross \cite{re:tesauro87}. Their analysis of the appropriate
amplitude equations for this case yields 
\begin{equation}
q_x^s = q_0 - \frac{\epsilon}{2 q_0^3}
\label{eq:qx_1}
\end{equation}
up to order $\epsilon$ for the selected wavenumber in domain A, and
$q_y^s=q_0$ up to order $\epsilon^{3/4}$ for domain B. They also conducted a
limited numerical study of this model to suggest that $q_y^s \simeq
q_0 - 0.12\epsilon$ (although this result is based on calculations
involving only two different values of $\epsilon$).

\begin{figure}
\includegraphics[width=2.8in]{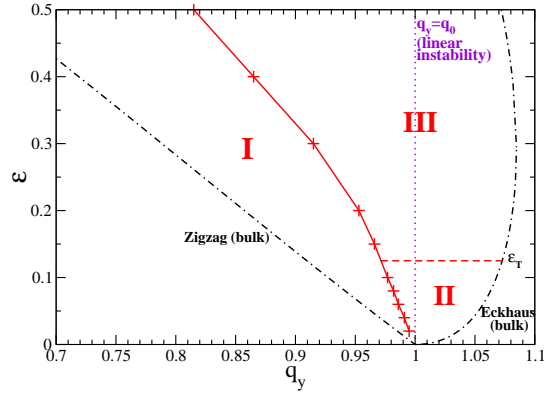}
\caption{Stability diagram for the non potential model 1 as a function of
  $\epsilon$ and $q_y$ (the wave number of the stripes perpendicular to
  the grain boundary). The plus symbols correspond to the unique
  wavevector $(q_x,q_y) = (q_x^s=q_0 - \epsilon/(2 q_0^3), q_y^s)$ for
  a stationary grain boundary configuration. Also shown (as dot-dashed
  lines) are stability boundaries for bulk Eckhaus and zig-zag
  instabilities. Three regimes for grain boundary decay are indicated,
  and also the critical wave number $q_0$ for linear instability.}
\label{fig:diagram}
\end{figure}

\begin{figure}
\includegraphics[width=2.8in]{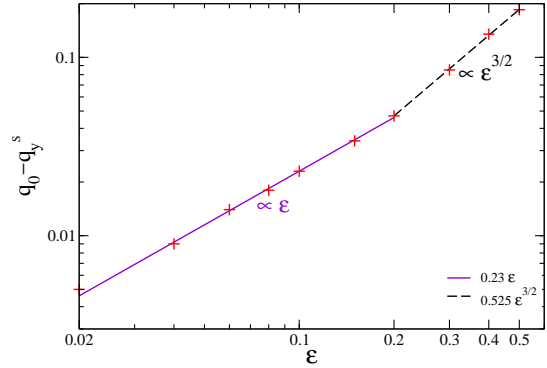}
\caption{Wave number $q_y^s$ for a stationary grain boundary (model 1)
  as a function of $\epsilon$. The figure shows two different types of
  dependency on $\epsilon$:  $q_0 - q_y^s \propto \epsilon$ (for small
  $\epsilon$) and 
  $ \propto \epsilon^{3/2}$ (for larger $\epsilon$).}
\label{fig:qxy_G}
\end{figure}

We have numerically calculated the selected wave numbers for the same
configuration but with much larger system sizes and longer integration times. Our
results for $q_x^s$ are in agreement with Eq. (\ref{eq:qx_1}), but the
calculated values of $q_y^s$ are different from those of previous studies, as
presented in Figs. \ref{fig:diagram} and \ref{fig:qxy_G}. The
deviation of $q_y^s$ from the critical wavenumber $q_0$ is seen to
obey two different power laws in $\epsilon$. For small
$\epsilon < \epsilon_0$, with $\epsilon_0$ around 0.2, we find
(Fig. \ref{fig:qxy_G}) 
\begin{equation}
q_y^s \simeq q_0 - 0.23 \epsilon,
\label{eq:qy1_1}
\end{equation}
while for larger $\epsilon$ we obtain,
\begin{equation}
q_y^s \simeq q_0 - 0.525 \epsilon^{3/2}.
\label{eq:qy2_1}
\end{equation} 
Note that although both results differ from the numerical calculations of
Ref. \onlinecite{re:tesauro87}, they are not inconsistent with an analysis
based on amplitude equations.

\subsection{Decay of the grain boundary configuration}
\label{subsec:regimes_1}

Given the results of Eqs. (\ref{eq:qx_1})--(\ref{eq:qy2_1}), and also our
numerical results in Fig. \ref{fig:diagram}, it is clear that the wave
number selected by a grain boundary is different from $q_0$, the most
unstable mode near onset. 
Thus the locally dominant wavenumber during the early stages of coarsening
could be significantly different from ($q_x^s,q_y^s$), and therefore it
appears to be of value to study the decay of grain boundaries under those
conditions. Of course, grain boundaries would not be fully formed at those
early stages, but we hope that studying unstable trajectories away from well
defined grain boundary configurations will illustrate differences in observed
morphologies between potential and non potential models.

We start by noting that in the configuration which we consider there is no
phase conservation in the $x$ direction. Hence if $q_x \neq q_x^s$ is chosen
initially, the bulk wave number of domain A can readjust to $q_x^s$ through
lamella formation or destruction near the boundary \cite{re:tesauro87}. We
will not consider this case. More complicated phenomena can occur for $q_y
\neq q_y^s$, and this is the focus of our study: In the calculations that
follow we always keep $q_x = q_x^s$ in domain A, while we explore a range of
values of $q_y$ in domain B. 

\begin{figure}
\includegraphics[width=6.4in]{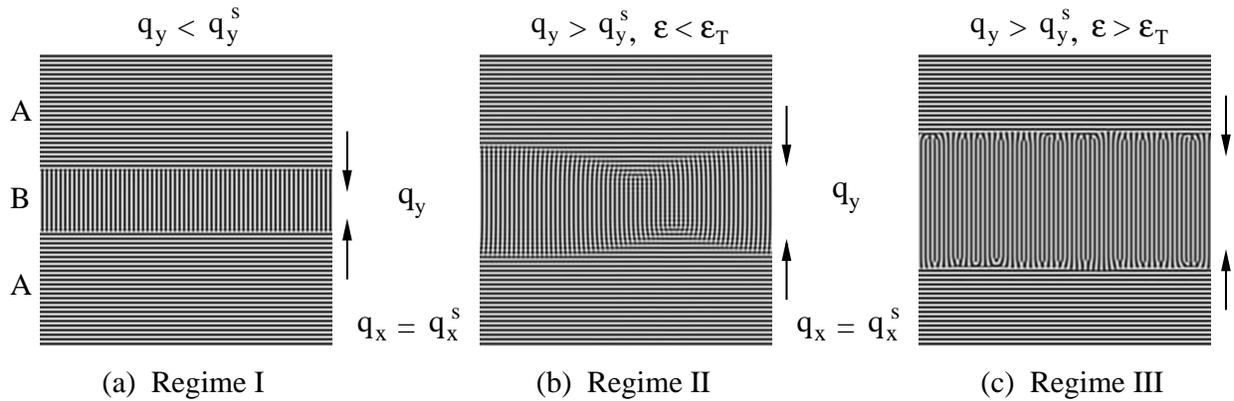}
\caption{Grain boundary decay in the three regimes shown in
  Fig. \ref{fig:diagram} for model 1, with $q_x=q_x^s$. The direction
  of boundary motion is indicated by arrows. (a) Regime I: $q_y<q_y^s$
  for all $\epsilon$; the example shown here corresponds to $q_y=0.85$,
  $\epsilon=0.3$ and at $t=1200$. (b) Regime II: $q_y>q_y^s$ and $\epsilon
  < \epsilon_T$; we have chosen $q_y=q_0=1$, $\epsilon=0.04$, and
  $t=10^4$. (c) Regime III: $q_y>q_y^s$ and $\epsilon
  > \epsilon_T$; the configuration shown here corresponds to $q_y=q_0=1$,
  $\epsilon=0.5$, and $t=100$.}
\label{fig:GBs1_regimes}
\end{figure}

As indicated in Fig. \ref{fig:diagram}, we find that our results can be
qualitatively classified according to the values of $q_{y}$ and $\epsilon$. In
regime I defined by $q_y<q_y^s$, the two grain boundaries move from domain A
towards domain B until they merge. Both boundaries remain approximately planar
during the process. Such a motion is found for all the values of $\epsilon$
investigated, with a particular example of $\epsilon=0.3$ and $q_y=0.85$
shown in Fig. \ref{fig:GBs1_regimes}a. 

On the other hand, a longitudinal distortion of the stripes in domain B around
the grain boundary region is observed if $q_y>q_y^s$. In addition, the
evolution following the appearance of the initial distortion is
qualitatively different depending on the value of $\epsilon$. In 
regime II of Fig. \ref{fig:diagram} with $\epsilon < \epsilon_T$ (where
the value of $\epsilon_T$ is around 0.1 to 0.2 as estimated from
our numerical solutions), long range undulations of diffuse grain boundaries
are observed, as shown in Fig. \ref{fig:GBs1_regimes}b. The stripes of
domain B are curved around the boundaries, the latter moving towards
each other at the expense of domain B until they finally merge. Despite the
undulations, the grain boundary configuration in regime II is
preserved during the process. In regime III on the other hand, the boundary is
sharp, and the longitudinal distortion leads to the formation of convex
disclinations in the boundary region (Fig. \ref{fig:GBs1_regimes}c for
$\epsilon > \epsilon_T$). The boundary itself also moves towards the bulk of
domain B as indicated by the arrows in the figure, resulting again in the
eventual disappearance of the grain boundary configuration. 

The longitudinal distortions that we observe in both regimes, including the
formation of an array of disclinations in regime III, are due to
wavenumber mismatch between the initial value $q_{y}$ and the stationary value
$q_{y}^{s} < q_{y}$. In regime II (Figs. \ref{fig:GBs1_regimes}b), a periodic
longitudinal modulation of stripes B appears, reminiscent of an Eckhaus
distortion that promotes wavenumber reduction. Since the velocity of the
boundary depends on the local wavenumber of stripes B ahead of it, a
modulation of the boundary position also results from this wavenumber variation.
In regime III, on the other hand, the boundary is sharp, and cross
rolls, dislocations and disclinations form at the edges of domain B
leading also to wavenumber reduction. The bulk of domains A and B,
however, remain largely unperturbed.

\begin{figure}
\includegraphics[width=4.3in]{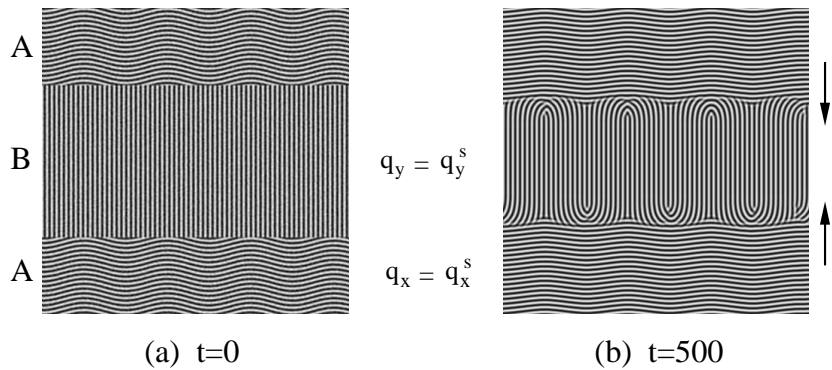}
\caption{Grain boundary decay (for model 1) due to a slow transverse modulation of
  domain A in an initial stationary configuration (with $q_x=q_x^s$ and
  $q_y=q_y^s$), with $\epsilon=0.5$, the modulation wavenumber
  $Q=q_x/16$, and the magnitude of the phase modulation $\delta=\lambda_0$.}
\label{fig:GBs1_modul}
\end{figure}

In order to verify that this spontaneous distortion originates in the
boundary region, we have studied the decay of the
configuration shown in Fig. \ref{fig:GBs1_modul}a in which 
slow transverse modulations of wavenumber $Q \ll q_x$ and phase
magnitude $\delta$ have been added to the initial condition of $q_x=q_x^s$
in domain A. In this case, $q_y=q_y^s$ in the bulk of domain B but
not at the grain boundaries. For large $\epsilon$
(regime III), the grain boundaries decay through the formation of
disclinations, as shown in Fig. \ref{fig:GBs1_modul}b. The same decay was
shown in Fig. \ref{fig:GBs1_regimes}c corresponding to the case $q_y \neq
q_y^s$ in the whole domain B.

A similar analysis can be carried out in regime II (small $\epsilon$), with
the same conclusion. We note, however, an interesting aspect of the decay as
compared to potential models. The grain boundary moves from the modulated
domain A towards the undistorted domain B, exactly the opposite of what is
expected for a potential model given our choice of wavenumbers.
With the consideration that wavenumbers $q_x$ and $q_y$ in the bulk of both
domains are equal to the stationary values $q_x^s$ and $q_y^s$ of the
corresponding potential model, the extra energy that is stored in the modulated 
stripes in domain A would lead to boundary motion in the opposite direction
\cite{re:boyer01}.

\section{Wavevector selection and grain boundary decay - Model 2}
\label{sec:model2}

The second non potential model studied, as described in Sec. \ref{sec:models}, 
includes the so called mean flows. They are known to have a significant effect on
the stability of a stripe configuration \cite{re:greenside85}, on defect dynamics
\cite{re:tesauro86,re:walter04}, and allow the appearance of spiral
defect chaos \cite{re:xi93}. We consider here the decay of the grain boundary
configuration studied above. We have chosen model parameter values that are
typical for fluid systems in Eqs. (\ref{eq:S-H})--(\ref{eq:mean_flow}):
$\sigma=1$, $c^2=2$, $g_m=10$, and $\gamma=\lambda_0/2$ for filtering.

We have first verified numerically that the unique wavenumbers for
a stationary grain boundary are $q_x^s = q_y^s = q_0$, the same as in the
potential Swift-Hohenberg model. This result follows from the observation that
the lowest order contribution of mean flow to the amplitude equations of a
grain boundary is ${\cal O}(\epsilon^{5/4})$ \cite{re:tesauro86}.

\begin{figure}
\includegraphics[width=4.3in]{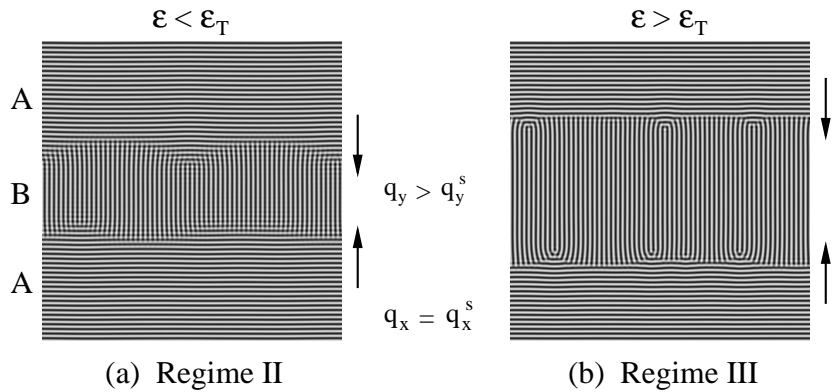}
\caption{Grain boundary decay for model 2 as a result of mean
  flows, with $g_m=10$, $\sigma=1$, $c^2=2$, and $q_x = q_x^s = q_0$. (a)
  Regime II: 
  $q_y>q_y^s$ and $\epsilon < \epsilon_T$; the parameters used in this case are
  $q_y=1.02$, $\epsilon=0.04$, and $t=4000$. (b) Regime III:
  $q_y>q_y^s$ and $\epsilon > \epsilon_T$; the configuration shown
  corresponds to $q_y=1.02$, $\epsilon=0.3$, and $t=600$.}
\label{fig:GBs2_regimes}
\end{figure}

The known stability diagram of a stripe configuration is similar to that shown
in Fig. \ref{fig:diagram} for model 1, except that $q_y^s=q_0=1$,
and the type and detailed location of the stability boundaries are
different (e.g., the Eckhaus boundary in Fig. \ref{fig:diagram} should
be replaced by the skewed-varicose boundary \cite{re:greenside85}). In terms
of our study of the decay of a grain boundary, we also find in this model  
three different regimes which are qualitatively similar to those of model
1. Rigid motion of the grain boundary is observed when $q_y < q_y^s$ for all
$\epsilon$, while a longitudinal distortion of the stripes in domain B appears
when  $q_y > q_y^s$. We can also qualitatively distinguish between regimes
II ($\epsilon < \epsilon_T$) and III ($\epsilon > \epsilon_T$), with two
typical examples given in Fig. \ref{fig:GBs2_regimes}.

As was the case in model 1, we observe long wavelength distortions of the
stripes in domain B in regime II (Fig. \ref{fig:GBs2_regimes}a), whereas
regime III leads to the formation and subsequent motion of disclinations, and
with them, the decay of the grain boundary configuration
(Fig. \ref{fig:GBs2_regimes}b). Disclinations also
appear as a result of local deviation of wavenumbers from $q_y^s$
in the grain boundary region.
This has been verified by adding slow transverse modulations on A stripes
while keeping ${\bm q}={\bm q^s}$ in the bulk of both domains, with
results that are qualitatively similar to those of Fig. \ref{fig:GBs1_modul}.

The driving force behind the appearance of longitudinal distortions, and with
them either long wavelength undulations or the formation of disclinations
is again local wavenumber reduction around the boundary. This is analogous to
what is observed in model 1, and has been already described in detail in
Sec. \ref{subsec:regimes_1}. However, once the undulations have been formed or
the defects nucleated, mean flows are effective in their subsequent motion.
This can be seen from the spatial distribution of the vorticity potential
$\zeta$, or the corresponding drift velocity ${\bm U}$ obtained in our
numerical solutions. In particular, we note that mean flows play a key
role in the motion of disclinations in regime III, a regime in which 
disclinations become pinned in their absence ($g_m \rightarrow 0$).

\section{Discussion and conclusions}
\label{sec:discussion}

The results described on the decay of a grain boundary configuration in
non potential 
models shows the existence of a range of parameters in which the decay leads to
the formation of defects (mostly dislocations and disclinations) in the
boundary region. This occurs for a configuration in which the stripes parallel
to the boundary plane have the selected wavenumber $q_{x}=q_{x}^{s}$, whereas
the stripes perpendicular to the boundary are under slight compression $q_{y}
> q_{y}^{s}$. Also, defects are only seen above a certain threshold $\epsilon
> \epsilon_{T}$. Below the threshold, undulations and motion of the boundary
follow, but no additional defects are produced.

\begin{figure}
\includegraphics[width=5.in]{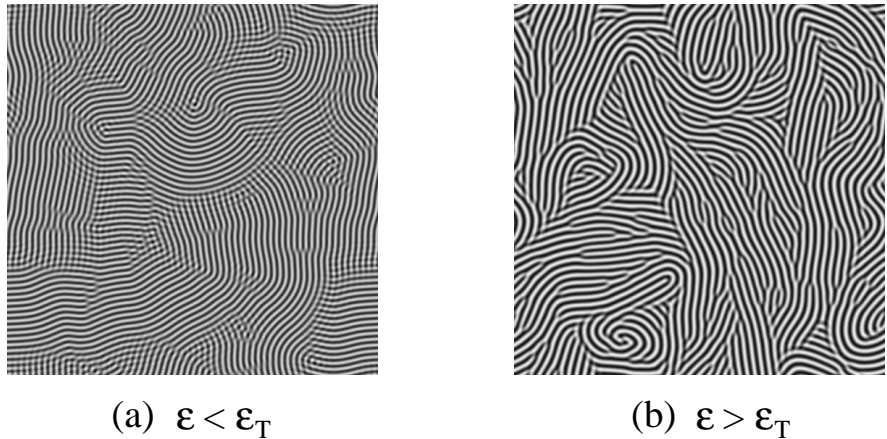}
\caption{Multidomain configurations for model 1 showing different qualitative
  coarsening behavior for (a) 
  $\epsilon < \epsilon_T$ (with $\epsilon=0.04$ and at $t=1000$) and
  (b) $\epsilon > \epsilon_T$ (with $\epsilon=0.5$ and at $t=500$). }
\label{fig:domains_1}
\end{figure}

These results can help interpret qualitative features of domain coarsening
behavior, both in potential and non potential models. In the case of
non potential model 1
(with a $|\nabla \psi|^2 \nabla^2 \psi$ nonlinearity), the
stationary wave numbers $q_x^s$ and $q_y^s$ are different from $q_0$,
the most unstable wave number (see Eqs. (\ref{eq:qx_1})--(\ref{eq:qy2_1}) and
Fig. \ref{fig:diagram}). Following the decay of an initially disordered
configuration ($\psi = 0$), and the appearance of locally ordered domains
with wavenumber approximately equal to $q_{0}$, grain boundaries would be
expected to decay according to the mechanism described for regimes II and III
depending on the value of $\epsilon$. Whether coarsening takes place in the
regime in which the decay of boundaries leads to the formation
of dislocations and disclinations (regime III), or to their motion (regime
II) one might anticipate qualitatively different late time domain
coarsening. Figure \ref{fig:domains_1} shows the results of a numerical
integration of model 1 from a initial random configuration of $\psi$ for two
particular values of $\epsilon$, both below and above $\epsilon_{T}$.
For $\epsilon < \epsilon_{T}$ (Fig. \ref{fig:domains_1}a), grain boundaries
are visible in the late time configuration as expected from the results shown
in Figs. \ref{fig:GBs1_regimes}b. 
Qualitatively different behavior is observed for $\epsilon > \epsilon_{T}$ due
to the early stage formation of dislocations or disclinations
around the emerging grain boundaries. Our results in this case are shown in
Fig. \ref{fig:domains_1}b, with the resulting disclination/dislocation
dominated morphology. These results are consistent with earlier observations
in  Ref. \onlinecite{re:cross95a}, and more recent research on the Boussinesq
equations in Ref. \onlinecite{re:paul04}. The average wave
number of the evolving configuration changes from $q_0$ at early times
following the instability, towards a long time value that is closer to the
stationary wave number of isolated dislocations. One anticipates that 
domain coarsening will be qualitatively different from the case of
small $\epsilon$, as suggested by the results of Ref. \onlinecite{re:cross95a} 
showing a complex scaling behavior at large $\epsilon$.

\begin{figure}
\includegraphics[width=5.in]{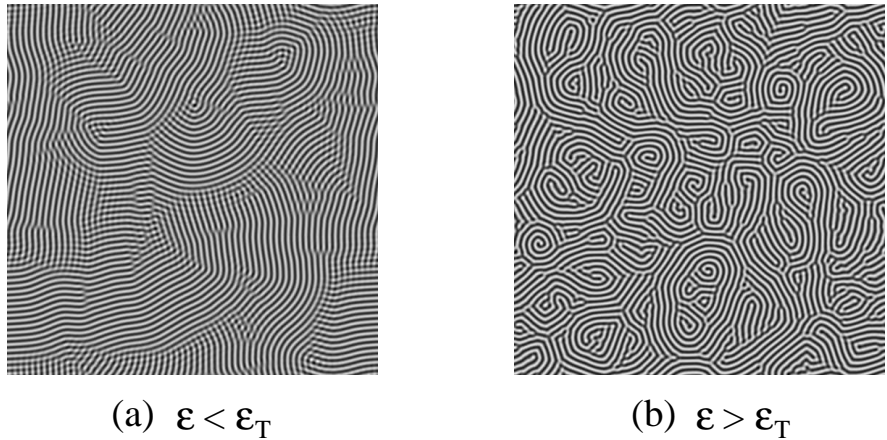}
\caption{Multidomain configurations of stripe phase in
  non potential model 2 (with mean flows), for (a)
  $\epsilon < \epsilon_T$ ($\epsilon=0.04$ and $t=1000$) and
  (b) $\epsilon > \epsilon_T$ ($\epsilon=0.5$ and $t=1000$). Other
  parameters are the same as those in Fig. \ref{fig:GBs2_regimes}: $g_m=10$,
  $\sigma=1$, and $c^2=2$.}
\label{fig:domains_2}
\end{figure}

In the case of model 2 which includes the coupling to mean
flows, the selected wavenumbers $q_x^s$ and $q_y^s$ approximately coincide with
the critical wavenumber for instability $q_0$, and therefore the effects
described above will be weaker, but not completely absent as local deviations
between the stripe wavenumber and the selected wavenumber still occur in
regions in which the stripes are curved. Near threshold ($\epsilon <
\epsilon_{T}$), the morphology of a transient stripe configuration is shown in 
Fig. \ref{fig:domains_2}a obtained by numerical solution of the governing
equations from a random initial condition. A stripe pattern qualitatively similar
to both model 1 and the potential Swift-Hohenberg model is found,
characterized by differently oriented domains separated by $90^{\circ}$ 
grain boundaries. On the other hand, when $\epsilon$ exceeds a certain threshold
$\epsilon_{T}$, grain boundaries are expected to be absent as
described in Sec. \ref{sec:model2}. The subsequent evolution of the system in
this case of is much more complicated resulting in persistent dynamics (spiral
defect chaos \cite{re:xi93}), as shown in Fig. \ref{fig:domains_2}b. 

In both non potential cases discussed in this paper we would expect
qualitatively different domain coarsening behavior depending on the interplay
between the critical wavenumber for instability and the selected wavenumber at
a grain boundary. This is in contrast with the potential Swift-Hohenberg model
in which grain boundary dynamics is known to dominate for small $\epsilon$ until
pinning effects at finite $\epsilon$ effectively arrest motion
\cite{re:boyer01b,re:boyer02}. In this model, the selected wavenumbers for
a stationary grain boundary configuration are
$q_x^s=q_y^s=q_0$ \cite{re:tesauro87}, with $q=q_0$ also marking the stability
boundary of the bulk zig-zag instability. Therefore a diagram analogous to
that of Fig. \ref{fig:diagram} would only have regimes II and III
characterizing grain boundary motion. In both $q_y > q_y^s$, longitudinal
distortions around the boundary can also be identified as a result
of local wavenumber reduction towards $q_y^s=q_0$. In analogy with the
non potential cases, in regime II (with $\epsilon < \epsilon_T$) diffuse
boundaries with long range undulations of B stripes
are observed. For $\epsilon > \epsilon_T$,
the formation of convex disclinations around the boundary can also be
seen in numerical calculations. However, despite
their emergence these disclinations become pinned along with the grain
boundaries, mainly due to the nonadiabatic effects arising from the
coupling between fast scales of base stripe patterns and slow scales
of the associated envelopes \cite{re:boyer02}. This results in a
glassy configuration with very slow dynamics.
Interestingly, this pinning phenomenon is only observed in the potential
Swift-Hohenberg model, but not in the two non potential models studied here. In
this latter case, grain boundaries quickly decay at large $\epsilon$ leaving
behind a domain morphology dominated by disclinations and dislocations
(Fig. \ref{fig:domains_1}b), or spiral defects (Fig. \ref{fig:domains_2}b).

In summary, we have studied wavenumber selection, motion, and
decay of grain boundaries in a stripe configuration based on a pair of
two-dimensional generalized Swift-Hohenberg model equations that are
non potential in nature. Three different regimes of grain boundary dynamics have
been identified, with longitudinal distortions emerging around the
grain boundary and manifesting themselves either as long range undulations
of the moving boundary or in the formation and subsequent motion of
defects such as convex disclinations. When an extended system is taken from an
initial state below threshold to $\epsilon > 0$, the initial decay of the
unstable state will lead to the formation of locally ordered domains with a
distribution of orientations. Domain interfaces will form in accordance with the
analysis given above, and therefore qualitative
differences in domain coarsening are expected depending on the value of $\epsilon$.

\begin{acknowledgments}
This research has been supported by the National Science Foundation
under grant DMR-0100903, and by NSERC Canada.
\end{acknowledgments}

\bibliography{../references}

\begin{thebibliography}{20}
\expandafter\ifx\csname natexlab\endcsname\relax\def\natexlab#1{#1}\fi
\expandafter\ifx\csname bibnamefont\endcsname\relax
  \def\bibnamefont#1{#1}\fi
\expandafter\ifx\csname bibfnamefont\endcsname\relax
  \def\bibfnamefont#1{#1}\fi
\expandafter\ifx\csname citenamefont\endcsname\relax
  \def\citenamefont#1{#1}\fi
\expandafter\ifx\csname url\endcsname\relax
  \def\url#1{\texttt{#1}}\fi
\expandafter\ifx\csname urlprefix\endcsname\relax\def\urlprefix{URL }\fi
\providecommand{\bibinfo}[2]{#2}
\providecommand{\eprint}[2][]{\url{#2}}

\bibitem[{\citenamefont{Seul and Andelman}(1995)}]{re:seul95}
\bibinfo{author}{\bibfnamefont{M.}~\bibnamefont{Seul}} \bibnamefont{and}
  \bibinfo{author}{\bibfnamefont{D.}~\bibnamefont{Andelman}},
  \bibinfo{journal}{Science} \textbf{\bibinfo{volume}{267}},
  \bibinfo{pages}{476} (\bibinfo{year}{1995}).

\bibitem[{\citenamefont{Bates and Fredrickson}(1999)}]{re:bates99}
\bibinfo{author}{\bibfnamefont{F.~S.} \bibnamefont{Bates}} \bibnamefont{and}
  \bibinfo{author}{\bibfnamefont{G.~H.} \bibnamefont{Fredrickson}},
  \bibinfo{journal}{Phys. Today} \textbf{\bibinfo{volume}{52}},
  \bibinfo{pages}{32} (\bibinfo{year}{1999}).

\bibitem[{\citenamefont{Cross and Hohenberg}(1993)}]{re:cross93}
\bibinfo{author}{\bibfnamefont{M.~C.} \bibnamefont{Cross}} \bibnamefont{and}
  \bibinfo{author}{\bibfnamefont{P.~C.} \bibnamefont{Hohenberg}},
  \bibinfo{journal}{Rev. Mod. Phys.} \textbf{\bibinfo{volume}{65}},
  \bibinfo{pages}{851} (\bibinfo{year}{1993}).

\bibitem[{\citenamefont{Purvis and Dennin}(2001)}]{re:purvis01}
\bibinfo{author}{\bibfnamefont{L.}~\bibnamefont{Purvis}} \bibnamefont{and}
  \bibinfo{author}{\bibfnamefont{M.}~\bibnamefont{Dennin}},
  \bibinfo{journal}{Phys. Rev. Lett.} \textbf{\bibinfo{volume}{86}},
  \bibinfo{pages}{5898} (\bibinfo{year}{2001}).

\bibitem[{\citenamefont{Elder et~al.}(1992)\citenamefont{Elder, {Vi\~nals}, and
  Grant}}]{re:elder92}
\bibinfo{author}{\bibfnamefont{K.}~\bibnamefont{Elder}},
  \bibinfo{author}{\bibfnamefont{J.}~\bibnamefont{{Vi\~nals}}},
  \bibnamefont{and} \bibinfo{author}{\bibfnamefont{M.}~\bibnamefont{Grant}},
  \bibinfo{journal}{Phys. Rev. Lett.} \textbf{\bibinfo{volume}{68}},
  \bibinfo{pages}{3024} (\bibinfo{year}{1992}).

\bibitem[{\citenamefont{Cross and Meiron}(1995)}]{re:cross95a}
\bibinfo{author}{\bibfnamefont{M.~C.} \bibnamefont{Cross}} \bibnamefont{and}
  \bibinfo{author}{\bibfnamefont{D.~I.} \bibnamefont{Meiron}},
  \bibinfo{journal}{Phys. Rev. Lett.} \textbf{\bibinfo{volume}{75}},
  \bibinfo{pages}{2152} (\bibinfo{year}{1995}).

\bibitem[{\citenamefont{Boyer and {Vi\~nals}}(2001{\natexlab{a}})}]{re:boyer01}
\bibinfo{author}{\bibfnamefont{D.}~\bibnamefont{Boyer}} \bibnamefont{and}
  \bibinfo{author}{\bibfnamefont{J.}~\bibnamefont{{Vi\~nals}}},
  \bibinfo{journal}{Phys. Rev. E} \textbf{\bibinfo{volume}{63}},
  \bibinfo{pages}{061704} (\bibinfo{year}{2001}{\natexlab{a}}).

\bibitem[{\citenamefont{Boyer and
  {Vi\~nals}}(2001{\natexlab{b}})}]{re:boyer01b}
\bibinfo{author}{\bibfnamefont{D.}~\bibnamefont{Boyer}} \bibnamefont{and}
  \bibinfo{author}{\bibfnamefont{J.}~\bibnamefont{{Vi\~nals}}},
  \bibinfo{journal}{Phys. Rev. E} \textbf{\bibinfo{volume}{64}},
  \bibinfo{pages}{050101(R)} (\bibinfo{year}{2001}{\natexlab{b}}).

\bibitem[{\citenamefont{Paul et~al.}(2004)\citenamefont{Paul, Chiam, Cross, and
  Fischer}}]{re:paul04}
\bibinfo{author}{\bibfnamefont{M.~R.} \bibnamefont{Paul}},
  \bibinfo{author}{\bibfnamefont{K.-H.} \bibnamefont{Chiam}},
  \bibinfo{author}{\bibfnamefont{M.~C.} \bibnamefont{Cross}}, \bibnamefont{and}
  \bibinfo{author}{\bibfnamefont{P.~F.} \bibnamefont{Fischer}},
  \bibinfo{journal}{Phys. Rev. Lett.} \textbf{\bibinfo{volume}{93}},
  \bibinfo{pages}{064503} (\bibinfo{year}{2004}).

\bibitem[{\citenamefont{Kamaga et~al.}(2004)\citenamefont{Kamaga, Ibrahim, and
  Dennin}}]{re:kamaga04}
\bibinfo{author}{\bibfnamefont{C.}~\bibnamefont{Kamaga}},
  \bibinfo{author}{\bibfnamefont{F.}~\bibnamefont{Ibrahim}}, \bibnamefont{and}
  \bibinfo{author}{\bibfnamefont{M.}~\bibnamefont{Dennin}},
  \bibinfo{journal}{Phys. Rev. E} \textbf{\bibinfo{volume}{69}},
  \bibinfo{pages}{066213} (\bibinfo{year}{2004}).

\bibitem[{\citenamefont{Harrison et~al.}(2000)\citenamefont{Harrison, Adamson,
  Cheng, Sebastian, Sethuraman, Huse, Register, and Chaikin}}]{re:harrison00b}
\bibinfo{author}{\bibfnamefont{C.}~\bibnamefont{Harrison}},
  \bibinfo{author}{\bibfnamefont{D.~A.} \bibnamefont{Adamson}},
  \bibinfo{author}{\bibfnamefont{Z.}~\bibnamefont{Cheng}},
  \bibinfo{author}{\bibfnamefont{J.~M.} \bibnamefont{Sebastian}},
  \bibinfo{author}{\bibfnamefont{S.}~\bibnamefont{Sethuraman}},
  \bibinfo{author}{\bibfnamefont{D.~A.} \bibnamefont{Huse}},
  \bibinfo{author}{\bibfnamefont{R.~A.} \bibnamefont{Register}},
  \bibnamefont{and} \bibinfo{author}{\bibfnamefont{P.~M.}
  \bibnamefont{Chaikin}}, \bibinfo{journal}{Science}
  \textbf{\bibinfo{volume}{290}}, \bibinfo{pages}{1558} (\bibinfo{year}{2000}).

\bibitem[{\citenamefont{Boyer and {Vi\~nals}}(2002)}]{re:boyer02}
\bibinfo{author}{\bibfnamefont{D.}~\bibnamefont{Boyer}} \bibnamefont{and}
  \bibinfo{author}{\bibfnamefont{J.}~\bibnamefont{{Vi\~nals}}},
  \bibinfo{journal}{Phys. Rev. E} \textbf{\bibinfo{volume}{65}},
  \bibinfo{pages}{046119} (\bibinfo{year}{2002}).

\bibitem[{\citenamefont{Greenside and Cross}(1985)}]{re:greenside85}
\bibinfo{author}{\bibfnamefont{H.~S.} \bibnamefont{Greenside}}
  \bibnamefont{and} \bibinfo{author}{\bibfnamefont{M.~C.} \bibnamefont{Cross}},
  \bibinfo{journal}{Phys. Rev. A} \textbf{\bibinfo{volume}{31}},
  \bibinfo{pages}{2492} (\bibinfo{year}{1985}).

\bibitem[{\citenamefont{Manneville}(1983)}]{re:manneville83}
\bibinfo{author}{\bibfnamefont{P.}~\bibnamefont{Manneville}},
  \bibinfo{journal}{J. Physique} \textbf{\bibinfo{volume}{44}},
  \bibinfo{pages}{759} (\bibinfo{year}{1983}).

\bibitem[{\citenamefont{Swift and Hohenberg}(1977)}]{re:swift77}
\bibinfo{author}{\bibfnamefont{J.}~\bibnamefont{Swift}} \bibnamefont{and}
  \bibinfo{author}{\bibfnamefont{P.~C.} \bibnamefont{Hohenberg}},
  \bibinfo{journal}{Phys. Rev. A} \textbf{\bibinfo{volume}{15}},
  \bibinfo{pages}{319} (\bibinfo{year}{1977}).

\bibitem[{\citenamefont{Tesauro and Cross}(1987)}]{re:tesauro87}
\bibinfo{author}{\bibfnamefont{G.}~\bibnamefont{Tesauro}} \bibnamefont{and}
  \bibinfo{author}{\bibfnamefont{M.~C.} \bibnamefont{Cross}},
  \bibinfo{journal}{Phil. Mag. A} \textbf{\bibinfo{volume}{56}},
  \bibinfo{pages}{703} (\bibinfo{year}{1987}).

\bibitem[{\citenamefont{Cross et~al.}(1994)\citenamefont{Cross, Meiron, and
  Tu}}]{re:cross94b}
\bibinfo{author}{\bibfnamefont{M.~C.} \bibnamefont{Cross}},
  \bibinfo{author}{\bibfnamefont{D.~I.} \bibnamefont{Meiron}},
  \bibnamefont{and} \bibinfo{author}{\bibfnamefont{Y.}~\bibnamefont{Tu}},
  \bibinfo{journal}{Chaos} \textbf{\bibinfo{volume}{4}}, \bibinfo{pages}{607}
  (\bibinfo{year}{1994}).

\bibitem[{\citenamefont{Tesauro and Cross}(1986)}]{re:tesauro86}
\bibinfo{author}{\bibfnamefont{G.}~\bibnamefont{Tesauro}} \bibnamefont{and}
  \bibinfo{author}{\bibfnamefont{M.~C.} \bibnamefont{Cross}},
  \bibinfo{journal}{Phys. Rev. A} \textbf{\bibinfo{volume}{34}},
  \bibinfo{pages}{1363} (\bibinfo{year}{1986}).

\bibitem[{\citenamefont{Walter et~al.}(2004)\citenamefont{Walter, Pesch, and
  Bodenschatz}}]{re:walter04}
\bibinfo{author}{\bibfnamefont{T.}~\bibnamefont{Walter}},
  \bibinfo{author}{\bibfnamefont{W.}~\bibnamefont{Pesch}}, \bibnamefont{and}
  \bibinfo{author}{\bibfnamefont{E.}~\bibnamefont{Bodenschatz}},
  \bibinfo{journal}{Chaos} \textbf{\bibinfo{volume}{14}}, \bibinfo{pages}{933}
  (\bibinfo{year}{2004}).

\bibitem[{\citenamefont{Xi et~al.}(1993)\citenamefont{Xi, Gunton, and
  {Vi\~nals}}}]{re:xi93}
\bibinfo{author}{\bibfnamefont{H.-W.} \bibnamefont{Xi}},
  \bibinfo{author}{\bibfnamefont{J.~D.} \bibnamefont{Gunton}},
  \bibnamefont{and}
  \bibinfo{author}{\bibfnamefont{J.}~\bibnamefont{{Vi\~nals}}},
  \bibinfo{journal}{Phys. Rev. Lett.} \textbf{\bibinfo{volume}{71}},
  \bibinfo{pages}{2030} (\bibinfo{year}{1993}).

\end{thebibliography}

\end{document}